\documentclass[useAMS, referee, usenatbib]{biom}

\usepackage{xcolor}
\usepackage{amsmath}
\usepackage{color, latexsym, epsfig, verbatim,  amsmath, amssymb, lscape,graphics,url}
\usepackage{algorithm}   % me
\usepackage{algpseudocode}
\usepackage{float}       % to fix figure position
\usepackage{xcolor}  % color for text 
\usepackage{graphicx, caption, subfig}  % include graph
\usepackage{booktabs}  % online table
\usepackage{natbib}
 \usepackage{setspace}
 \usepackage{amsmath,amssymb}
\usepackage{appendix}
\usepackage{breqn}   % equation 
 
\def\bSig\mathbf{\Sigma}

    \def\bbeta{\mbox{\boldmath $\beta$}}

    \def\bgamma{\mbox{\boldmath $\gamma$}} 
     \def\bepsilon{\mbox{\boldmath $\epsilon$}}

    \def\bH{{\bf H}}

    \def\bW{{\bf W}}
    \def\bZ{{\bf Z}}

    \def\bI{{\bf I}}

    \def\bO{{\bf O}}

\DeclareMathOperator*{\argmin}{arg\,min}

\title{Unlocking the Power of Multi-institutional Data: Integrating and Harmonizing Genomic Data Across Institutions}
\author{Yuan Chen$^{1,*}$\email{cheny19@mskcc.org}, 
Ronglai Shen$^{1}$, Xiwen Feng$^{2}$, and Katherine Panageas$^{1}$ \\
$^{1}$Department of Epidemiology \& Biostatistics, Memorial Sloan Kettering Cancer Center, \\ New York, New York, U.S.A.  \\
$^{2}$Department of Biostatistics, University of Michigan,
Ann Arbor, Michigan, U.S.A.}

\begin{document}

%  This will produce the submission and review information that appears
%  right after the reference section.  Of course, it will be unknown when
%  you submit your paper, so you can either leave this out or put in 
%  sample dates (these will have no effect on the fate of your paper in the
%  review process!)

%\date{{\it Received November} 2023. {\it Revised February} 2024.  {\it
%Accepted March} 2024.}

%  These options will count the number of pages and provide volume
%  and date information in the upper left hand corner of the top of the 
%  first page as in published papers.  The \pagerange command will only
%  work if you place the command \label{firstpage} near the beginning
%  of the document and \label{lastpage} at the end of the document, as we
%  have done in this template.

%  Again, putting a volume number and date is for your own amusement and
%  has no bearing on what actually happens to your paper!  

%\pagerange{\pageref{firstpage}--\pageref{lastpage}} 
%\volume{64}
%\pubyear{2023}
%\artmonth{November}

%  The \doi command is where the DOI for your paper would be placed should it
%  be published.  Again, if you make one up and stick it here, it means 
%  nothing!

%\doi{10.1111/j.1541-0420.2005.00454.x}

%  This label and the label ``lastpage'' are used by the \pagerange
%  command above to give the page range for the article.  You may have 
%  to process the document twice to get this to match up with what you 
%  expect.  When using the referee option, this will not count the pages
%  with tables and figures.  

\label{firstpage}

%  put the summary for your paper here
%  225 word limit for abstract

\begin{abstract}

Cancer is a complex disease driven by genomic alterations, and tumor sequencing is becoming a mainstay of clinical care for cancer patients. The emergence of multi-institution sequencing data presents a powerful resource for learning real-world evidence to enhance precision oncology. GENIE BPC, led by American Association for Cancer Research, establishes a unique database linking genomic data with clinical information for patients treated at multiple cancer centers. However, leveraging sequencing data from multiple institutions presents significant challenges. Variability in gene panels can lead to loss of information when analyses focus on genes common across panels. Additionally, differences in sequencing techniques and patient heterogeneity across institutions add complexity. High data dimensionality, sparse gene mutation patterns, and weak signals at the individual gene level further complicate matters. Motivated by these real-world challenges, we introduce the Bridge model. It uses a quantile-matched latent variable approach to derive integrated features to preserve information beyond common genes and maximize the utilization of all available data, while leveraging information sharing to enhance both learning efficiency and the model's capacity to generalize. By extracting harmonized and noise-reduced lower-dimensional latent variables, the true mutation pattern unique to each individual is captured. We assess model’s performance and parameter estimation through extensive simulation studies. The extracted latent features from the Bridge model consistently excel in predicting patient survival across six cancer types in GENIE BPC data.

%We develop a model estimation procedure to prevent overfitting, ensuring robust applicability to new patient populations. 

%The AACR GENIE-BPC project is a multi-institution collaboration to generate a real-world, observational database linking genomic data with clinical information across institutions, providing a powerful resource for addressing those scientific and clinical questions. 

\end{abstract}

%  Please place your key words in alphabetical order, separated
%  by semicolons, with the first letter of the first word capitalized,
%  and a period at the end of the list.
%


%  As usual, the \maketitle command creates the title and author/affiliations
%  display 

\maketitle

%  If you are using the referee option, a new page, numbered page 1, will
%  start after the summary and keywords.  The page numbers thus count the
%  number of pages of your manuscript in the preferred submission style.
%  Remember, ``Normally, regular papers exceeding 25 pages and Reader Reaction 
%  papers exceeding 12 pages in (the preferred style) will be returned to 
%  the authors without review. The page limit includes acknowledgements, 
%  references, and appendices, but not tables and figures. The page count does 
%  not include the title page and abstract. A maximum of six (6) tables or 
%  figures combined is often required.''

%  You may now place the substance of your manuscript here.  Please use
%  the \section, \subsection, etc commands as described in the user guide.
%  Please use \label and \ref commands to cross-reference sections, equations,
%  tables, figures, etc.
%
%  Please DO NOT attempt to reformat the style of equation numbering!
%  For that matter, please do not attempt to redefine anything!

\section{Introduction} \label{intro}

Cancer is a multifaceted and heterogeneous disease in which genomic alterations can profoundly impact its initiation, treatment, and progression. The widespread adoption of genomic profiling in various cancer types has been facilitated by both commercial endeavors and initiatives driven by cancer research centers \citep{berger2018emerging, chakravarty2021clinical, pugh2022aacr}. \textcolor{black}{One leading effort for a multi-institutional cancer data repository was led by American Association for Cancer Research (AACR). AACR Project GENIE \citep{aacr2017aacr} creates a publicly accessible cancer registry of real-world genomic data assembled through data sharing between 19 leading international cancer centers. GENIE BPC (Biopharma Collaborative) \citep{de2023analysis} further links the multi-institutional clinical-grade genomic data with electronic health record-based clinical information for tens of thousands of cancer patients.}
The emergence of such real-world data presents exceptional opportunities that extend beyond single institution research studies. The integration of data across multiple institutions enables aggregation of larger datasets and more diverse patient representation. Consequently, this unlocks significant potential for investigating rare cancer types, specific patient subgroups, and uncommon genetic variants.

\begin{figure}[!t]
        \vskip -0.2 in
          \centering    
  \centering
    \includegraphics[width=1 \linewidth]{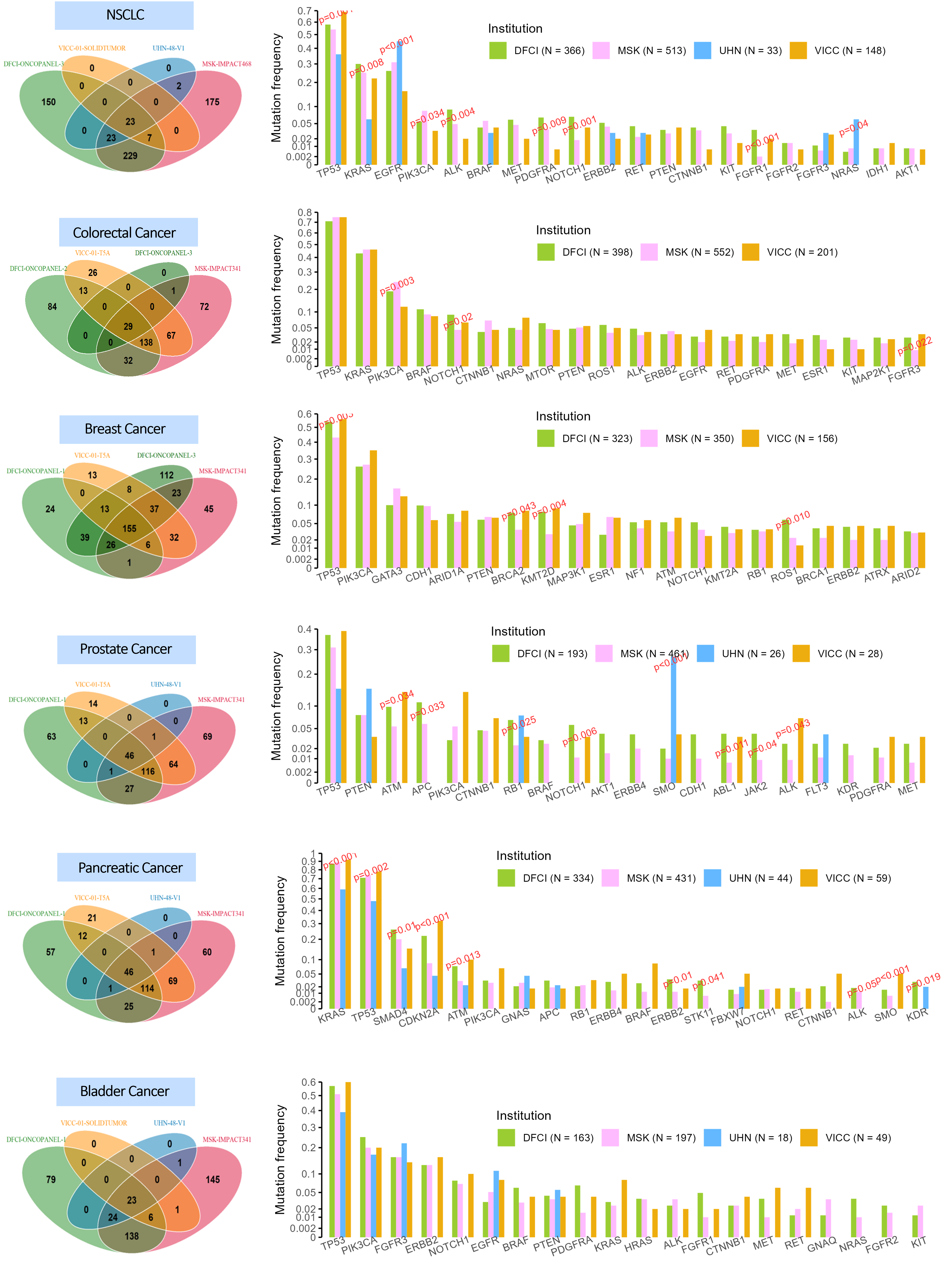}
    % \vskip -0.1 in
\vskip -0.05 in
\caption{Left panel: Venn diagrams illustrating the limited common genes (among genes with at least one observed mutation) covered by selected sequencing panels across institutions. Right panel: Mutation frequency by institution (indicated by the color of the bar) for the top 20 common genes of each cancer type in GENIE BPC.}
     % \vskip -0.21 in
     \label{venn_bar_6cancer}
\end{figure}

While the prospect of conducting genomic investigations across multiple institutions is promising, it poses several significant challenges. The primary obstacle stems from the genomic profiling efforts within each institution, leading to substantial variations in the design of targeted sequencing panels and assay techniques across institutions \citep{ harismendy2009evaluation, aacr2017aacr, fancello2019tumor}. For example, GENIE BPC comprises over 600 genes across all panels, but less than 50 genes are sequenced within all panels. Figure \ref{venn_bar_6cancer} provides an illustration of the scarcity of shared genes for each cancer type in GENIE BPC. For a comprehensive understanding of each patient’s genomic profile, it is imperative to leverage all available information for each individual rather than constrain the analysis to the common genes sequenced across all patients.

\begin{figure}[!t]
        \vskip -0.2 in
          \centering    
  \centering
    \includegraphics[width=1 \linewidth]{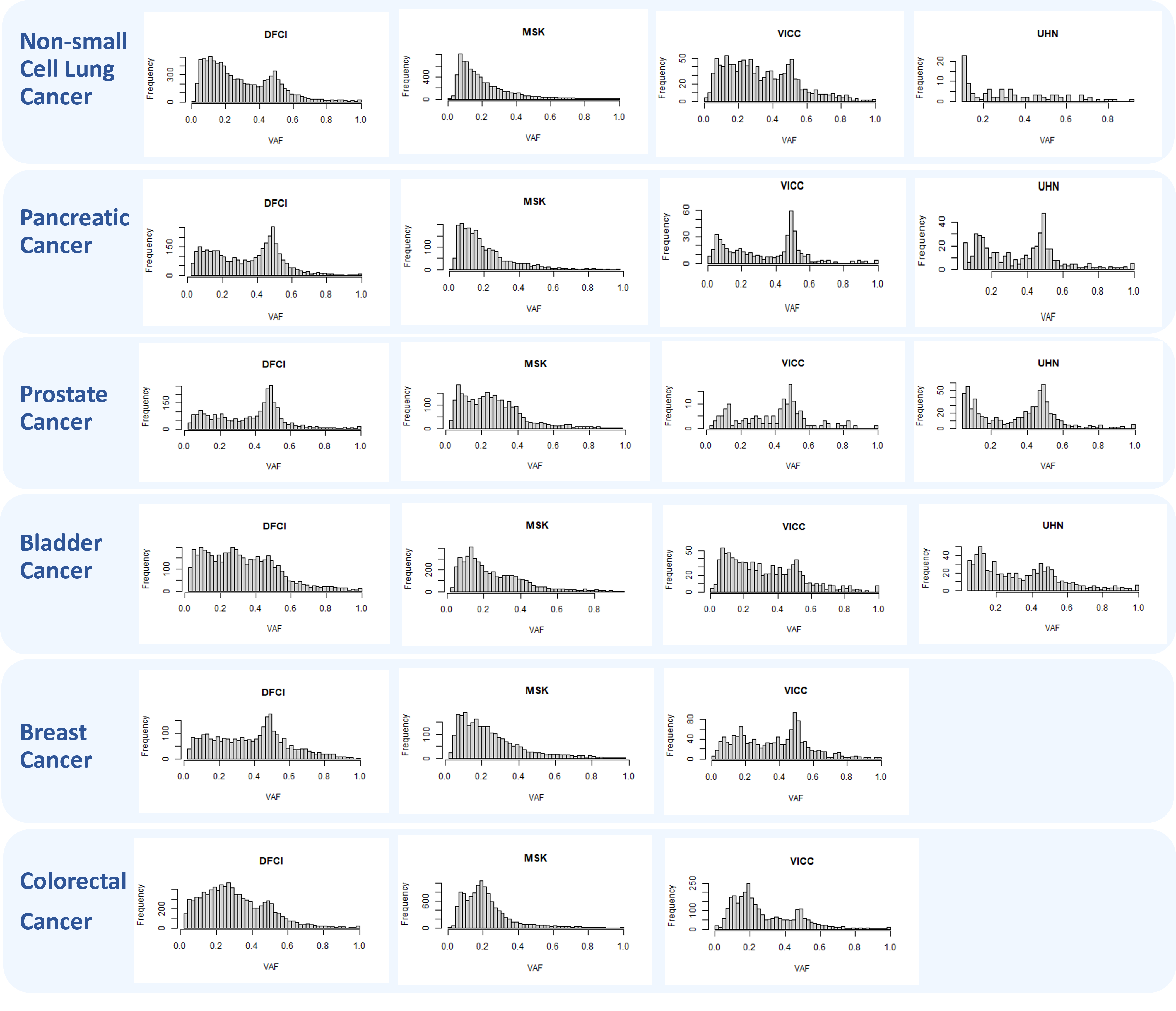}
    % \vskip -0.1 in
\vskip -0.05 in
    \caption{Distribution of variant allele frequency (VAF) by institution for each cancer type in GENIE BPC. VAF is defined as the number of variant alleles divided by the total number of sequenced alleles.}
     % \vskip -0.21 in
     \label{VAF_6cancer}
\end{figure}

Furthermore, differences in sequencing techniques can result in systematic discrepancies in mutation detection \citep{shi2018reliability, nakamura2011sequence, garofalo2016impact}, while patient heterogeneity introduces an additional layer of complexity to the observed alteration heterogeneity across institutions. In GENIE BPC, variations in sequencing techniques include gene panel coverage, required tumor purity, sequence depth, and mutation calling algorithms \citep{aacr2017aacr}. For instance, Memorial Sloan Kettering Cancer Center (MSK) employed somatic mutation calling based on matched tumor-normal samples, whereas Dana Farber Cancer Institute (DFCI) and Vanderbilt-Ingram Cancer Center (VICC) utilized tumor-only sequencing. Tumor-only sequencing presents unique challenges in distinguishing somatic mutations from germline variants without the availability of a matched normal sample from the same individual, and the false positive rate of somatic variant calling can be substantial in tumor-only sequencing assays \citep{shi2018reliability, asmann2021inflation, garofalo2016impact}. In GENIE BPC, distinct distributions of variant allele frequency (VAF) across institutions are observed in Figure \ref{VAF_6cancer}, and a second mode centered around the value of 0.5 for DFCI and VICC is likely attributed to germline mutations contaminating somatic mutation calls in tumor-only sequencing \citep{sun2018computational, smith2016somvarius}. Overlooking discrepancies in sequencing platforms can impede our ability to reveal clinically relevant genomic insights, as the true signals may be weakened or obscured by inconsistent measurements. Furthermore, we observed significant cohort heterogeneities across institutions regarding the clinical factors. For example, for non-small cell lung cancer (NSCLC), UHN exhibited a notably younger patient population, while VICC had fewer female patients and a higher prevalence of current smokers; for prostate cancer, MSK had a higher proportion of metastatic cancer at diagnosis, while patients at UHN exhibited the highest Gleason score. Examining the individual genes in Figure \ref{venn_bar_6cancer} shows significant variations in the difference of observed mutation frequency between institutions across different genes, which could arise from both technical variations or true patient heterogeneities.

Additionally, navigating the clinical implications within the high-dimensional genomic feature space presents an additional challenge. The high dimensionality of the data not only constrains statistical power but can also introduce noise to statistical models. Among the vast array of gene mutations, some or perhaps many, may lack meaningful associations or offer limited insights into clinical outcomes \citep{andreyev2001kirsten}. The risk of overfitting arises in such high-dimensional, low-signal data, resulting in poor generalization to new data.

Without methodologically substantive efforts to address all aforementioned challenges, the construction of clinico-genomic real-world datasets such as GENIE BPC is likely to have limited utility.
Latent variables have emerged as a powerful tool for integrating data from various data sources or data domains. It has found wide applications in psychology and social sciences for studying latent traits\citep{bollen2002latent, keyes2005mental}. 
The study by \cite{chen2021learning} highlighted its superiority over traditional methods in tailoring mental disorder treatment.
%In the study by \cite{chen2021learning}, machine learning models with latent variables were employed to integrate different questionnaire responses used for evaluating mental disorders, effectively capturing the nuanced nature of domain-specific impairments. They show that the extracted latent variables outperformed traditional predictors and outcome variables for treatment tailoring. 
In genomics, it’s been successfully adopted for integrating multi-omics data, reducing dimensionality, and enhancing clinical utility \citep{shen2012integrative, lock2013joint, gaynanova2019structural}.
%They have demonstrated the advantage of employing latent variable modeling approach for integrative analysis and dimensionality reduction, enabling the extraction of meaningful signals from high-dimensional feature sets.
%For example, \cite{shen2012integrative} developed an integrative clustering approach that incorporates multiple genomic data types in building a single cluster assignment. \cite{lock2013joint} and \cite{gaynanova2019structural} introduced integrative decomposition models that capture shared and individual structures of each data type. They have demonstrated the advantage of utilizing low-dimensional latent variables as a powerful method for integrative analysis and dimensionality reduction, enabling the extraction of meaningful signals from high-dimensional feature sets. 
% However, all these methods were designed for studies where diverse data modalities are gathered from the same set of patients. They are inadequate for the integration of genomic data from subjects across different institutions, given the unique challenges stemming from disparities in sequencing panels and techniques.
On the other hand, joint analysis of multi-study genomic data has yielded fruitful results \citep{riester2014risk, aggarwal2018clinical}. A recent study \citep{de2021bayesian} proposed multi-study factor analysis to integrate gene expression data from multiple breast cancer studies incorporating both common and study-specific factors. It demonstrated that the common latent factors derived from multi-study data hold greater clinical significance. While the study-specific latent factors may partly account for the technical variations between sequencing panels, they may also capture real variations stemming from the diverse study populations. Additionally, the issue of missing data caused by differences in the composition of gene panels, a common challenge in genomic datasets from targeted sequencing across multiple institutions, was not tackled.

Therefore, we introduce a statistical model, referred to as ``Bridge", explicitly designed to bridge and harmonize multi-institutional genomic data. This model encompasses genes commonly shared among institutions as well as those unique to specific institutions, aiming to ensure comprehensive and impartial information preservation. In the Bridge model, we address the unique challenges posed by significant panel differences and mitigate the inconsistencies in mutation detection measurements while preserving the true biological variations of individuals. Consequently, we derive a set of harmonized lower-dimensional latent variables that effectively capture the intrinsic mutation pattern of each subject. These variables are free from systematic biases, eliminate noise at the individual gene level, and ensure maximal preservation of each patient's genomic information. They facilitate effortless integration into subsequent analyses, which has the potential to offer valuable insights into patient heterogeneity, inform individual clinical outcomes, elucidate treatment mechanisms, and advance the field of precision oncology.

% They can readily facilitate seamless integration into subsequent analyses, providing valuable insights for patient heterogeneity, to inform patient clinical outcomes, study treatment mechanisms and promote precision oncology.

To structure our paper, we arrange the content as follows. We introduce the Bridge model and provide details on model estimation and generalization in Section \ref{method}. In Section \ref{simulation}, we present extensive simulation studies, examining the performance of the Bridge model. This includes comparative analysis against alternative models and evaluations of parameter estimations. In Section \ref{real_data}, we apply the Bridge model to all six cancer types in GENIE BPC data and demonstrate its superiority over alternative methods. Lastly, we provide a discussion in Section \ref{discussion} to contextualize our finding and their broader implications.

\vspace{-0.2in}
\section{Method} \label{method}
\subsection{Model framework}
Assume we have tumor sequencing data from $n$ subjects collected from $M$ different institutions. For subject $i = 1,..., n$, let $Y_{ij} \in \{0, 1 \}$ denote whether mutation was detected in gene $j$ $(j = 1,..., p)$, and $C_i \in \{1,...,M\}$ be the index of the institution where subject $i$'s sequencing was performed. There are two important considerations for $Y_{ij}$. Firstly, gene $j$ may not undergo sequencing at all institutions, resulting in the absence of observations for $Y_{ij}$ for certain subjects. Secondly, detecting a mutation $Y_{ij}$ may not always signify a genuine mutation, and similarly, a negative detection does not necessarily imply a conclusive absence of a variant, due to the variations in sequencing techniques as discussed in Section \ref{intro}.

We further let $V$ be a vector of patient baseline characteristics, e.g., age, gender, diagnosis stage, and cancer histology, potentially explaining part of the heterogeneity in genomic mutations among subjects, and we let $V_i$ represent the value of $V$ for subject $i$. More importantly, we employ a set of lower-dimensional subject-specific latent variables, represented by $Z \in \mathbb{R}^{K}$, to capture individual-level deviations from the population means that are not explained by the common covariate effects from $V$. Similarly, we let $Z_i$ denote the value of $Z$ for subject $i$. \textcolor{black}{These latent variables $Z_i$ serve as effective dimension reduction for the high-dimensional genomic features of $Y_{ij}$ and represent the underlying unique mutation patterns specific to each individual. }
Connecting all the variables together, we propose the following model for the observed mutation status $Y_{ij}$:
\begin{equation} \label{y_prob}
\mathbb P(Y_{ij} = 1 \mid \mathrm{C}_i = m, Z_i, V_i) = \sigma \Big( \beta_{jm} + \bW_j Z_i + \bgamma_j^T V_{i} \Big),
\end{equation}
where $\sigma: \mathbb{R} \rightarrow (0, 1)$ is the link function, for example, the sigmoid function $\sigma(x) = \frac{1}{{1 + e^{-x}}}$. In model (\ref{y_prob}), the association between patient characteristics $V$ and mutation in gene $j$ are captured by the parameters $\bgamma_j \in \mathbb{R}^{L}$, and the association between $Z_i$ and $Y_{ij}$ is captured by $\bW_j \in \mathbb{R}^{1 \times K}$, the $j$-th row of matrix  $\bW \in \mathbb{R}^{p \times K}$. Additionally, we introduce gene and institution-specific parameters $\{\beta_{jm}, j \in I_m, m=1,...,M\},$ where $I_m$ is the set of genes that are sequenced at institution $m$. The $\{\beta_{jm}\}$ terms are designed to account for the disparities in mutation detection that arise due to variations in sequencing platforms. %after accounting for the heterogeneity stemming from intrinsic individual differences from $V_i$ and the subject-specific mutation patterns in $Z_i$ under the shared latent structure $\bW$.

To this end, model (\ref{y_prob}) identifies the heterogeneity in the detected mutations through three components. Our primary focus lies on the latent variables, $Z_i$, which serve as a compact, denoised, and harmonized representation of subject-specific mutation patterns after effectively addressing the impact of panel differences without overcorrecting for heterogeneity resulting from patient characteristics. Consequently, these latent variables, which are not affected by missing data, present as appealing genomic features that can be seamlessly incorporated into subsequent analyses. %They hold the potential to elucidate patient heterogeneity, improve the accuracy of the prediction of patient outcomes, and contribute to the advancement of personalized medicine and treatment strategies.
Furthermore, $Z_i$ will be inferred by leveraging comprehensive genomic data sequenced for subject $i$, including both shared and institution-specific genes, to ensure maximal information preservation. Details on this will be provided in Section \ref{model_estimation}.

% To learn the model parameters and the subject-specific latent variables $\bZ$, our objective is to minimize the cross-entropy loss between the observed $Y_{ij}$ and its modeled probability. The cross-entropy loss corresponds to the negative log-likelihood for binary variables. The loss function takes the following form

% \begin{dmath*}
% %\textrm{min} %_{\bW, \bZ, \beta_{jm}}
% - \sum_{i=1}^n \sum_{j \in I_{ij}}
% \left \{ Y_{ij} \log \left(\sigma \left(\textcolor{black}{ \sum_{m=1}^M \beta_{jm} I(C_i = m)} + \bW_j Z_i + \bgamma_j^T V_{i} \right)\right) +  
% (1 - Y_{ij}) \log \left(\sigma \left (\textcolor{black}{\sum_{m=1}^M \beta_{jm} I(C_i = m)} + \bW_j Z_i + \bgamma_j^T V_{i} \right) \right) \right \}, \\
% %\textrm{s.t. } {\bW^T \bW = \bI,  \quad \sum_i \bZ_i I(C_i = m) = \boldmath{0} \quad \forall m},
% \end{dmath*}
% where $I_{ij}$ is the indices for the set of genes sequenced for subject $i$. $I_{ij}$ usually differs across institutions due to sequencing panel composition differences. Our model also allows $I_{ij}$ to differ across individuals to accommodate potential missing data issues.

To preserve greater flexibility in model forms, we avoid imposing any distribution assumptions on the latent variables $Z$. Instead, we assume that once we account for cohort differences and patient intrinsic heterogeneity, the latent traits $Z$ for patients should exhibit a similar distribution across institutions in the absence of sequencing panel differences.
Note that the parameters $\{\beta_{jm}\}$ introduced in model (\ref{y_prob}) address systematic sequencing differences on the mean level. However, to further align the distribution of the non-parametric $Z$ across institutions, we introduce regularization terms that focus on the differences in quantiles for each dimension of $Z$ across institutions.
Quantile-based alignment, as demonstrated in the literature, has found successful applications in standardizing and normalizing gene expression data across various platforms or data sources \citep{hansen2012removing, zyprych2015impact, valikangas2018systematic}.
In our approach, we encourage alignment of the quantiles for the latent variables $Z$ across institutions. Specifically, we let 
$Q(Z; p) =  \inf \left\{z \in \mathbb {R} :p\leq \widehat F_Z(z)\right\} ,
$
denote the empirical quantile function of $Z$, 
where $\widehat F_Z(z) = \frac{1}{n} \sum_{i=1}^n I (Z_i \le z) $ is the empirical cumulative distribution function of $Z$, and $p$ is the probability level. 
Furthermore, we let $\bZ := [Z_1, Z_2, ..., Z_n]$ represent the $K \times n$ matrix containing the $Z$ values for the $n$ subjects. We also specify $\mathbf{Z}^{(m)}$ as the subset of columns in $\mathbf{Z}$ corresponding to the patients who underwent sequencing at institution $m$, with $\mathbf{Z}_k^{(m)}$ representing the $k$-th row of $\mathbf{Z}^{(m)}$. Then the regularization terms can be constructed as
\begin{equation} \label{z_penalty}
\sum_{s=1}^S \sum_{k=1}^K \sum_{m=2}^M 
\left \{Q(\bZ_{k}^{(m)}; p_s) - Q(\bZ_{k}^{(m-1)}; p_s)
\right \}^2.
\end{equation}
These terms compute quantile differences between pairs of institutions for each dimension of $Z$. We will incorporate these regularization terms in the objective function to encourage the alignment of the $Z$ distributions considering the prespecified set of $S$ quantiles.
More details are provided in Section \ref{model_estimation_objective}. %This approach, complementing the corrections made at the mean level, will further facilitate mitigating the measurement inconsistencies across institutions, especially considering the non-parametric nature of $Z$.

\subsection{Model estimation} \label{model_estimation}

\subsubsection{Objective function} \label{model_estimation_objective}
To estimate the parameters in model (\ref{y_prob}) and the subject-specific latent variables $Z_i$'s, our primary objective is to minimize the cross-entropy loss between the observed $Y_{ij}$ and its modeled probability. This loss corresponds to the negative log-likelihood for binary variables. Additionally, we will incorporate the regularization terms specified in (\ref{z_penalty}) and two sets of constraints for model parameter identifiability. The resulting objective function is formulated as follows:

\begin{align}
\min_{{\beta_{jm}, \bW_j, \gamma_j, Z_i,\ \forall j, m, i}}  & - \frac{1}{n} \sum_{i=1}^n \frac{1}{|I_{i}|} \sum_{j \in I_{i}} \Bigg( Y_{ij} \log\bigg[\sigma\Big\{\sum_{m=1}^M \beta_{jm} I(C_i = m) + \bW_j Z_i + \bgamma_j^T V_{i}\Big\}\bigg] \notag \\
&  + (1 - Y_{ij}) \log\bigg[1 - \sigma\Big\{\sum_{m=1}^M \beta_{jm} I(C_i = m) + \bW_j Z_i + \bgamma_j^T V_{i}\Big\}\bigg] \Bigg)  \notag \\
& + \lambda \sum_{s=1}^S \sum_{k=1}^K \sum_{m=2}^M \bigg\{Q(\bZ_{k}^{(m)}; p_s) - Q(\bZ_{k}^{(m-1)}; p_s)\bigg\}^2  \notag \\
\textrm{s.t. } & {\bW^T \bW = \bI, \quad \sum_i Z_{ik} I(C_i = m) = \boldmath{0}, \quad \forall k = 1,...,K, \quad m = 1,...,M, } \label{objective}
\end{align}
where $I_{i}$ represents the indices corresponding to the set of genes sequenced for subject $i$, $|I_{i}|$ denotes the cardinality of this set, and $\lambda$ serves as a tuning parameter that governs the strength of the regularization terms. 
We note that $I_{i}$ usually differs across institutions due to sequencing panel composition differences, and here, we further allow it to differ across individuals to accommodate potential missing data on the individual level. Therefore, the latent representation $Z_i$ for each subject is inferred from their comprehensive genomic data to fully capture their unique mutation patterns. The learning of $\bW_j$ utilizes all subjects with data for gene $j$. Consequently, the model leverages all available data while borrowing information for increased learning efficiency and generalizability.

\subsubsection{Identifiability}

To make the gene- and institution-specific parameters $\{\beta_{jm}\}$ identifiable, we pose a set of constraints for $Z_i$'s as
\begin{align} \label{z_constraint}
    \sum_{i=1}^n Z_{ik} I(C_i = m) = \boldmath{0}, \quad \forall k = 1,...,K, \quad m = 1,...,M, 
\end{align} 
where $Z_{ik}$ denotes the $k$'th latent variable in $Z_{i}$. This essentially requires that within each institution $m$, each dimension of $Z$ has mean of $0$. It can be easily shown that if $(\widetilde \beta_{jm}, \widetilde \bW, \widetilde Z_i)$ is another parameterization of $(\beta_{jm}, \bW, Z_i)$ with $\widetilde Z_i$ satisfying the same constraint in (\ref{z_constraint}), then 
$ \beta_{jm} - \widetilde \beta_{jm} = 1/n_m \sum_{i=1}^n (\widetilde \bW_j \widetilde \bZ_i - \bW_j \bZ_i) I(C_i = m) = 1/n_m \{ \widetilde \bW_j \sum_{i=1}^n \widetilde \bZ_i I(C_i = m) - \bW_j \sum_i \bZ_i I(C_i = m)\} = 0,
$
where $n_m$ is the number of patients at institution $m$.

Additionally, we require columns of the loading matrix $\bW$ to be orthonormal, i.e., $\bW^T \bW = \bI$ to reduce the multicollinearity and to make the scale of ${Z_i}$ identifiable. Note that $\bW$ and $\bZ$ are not separately identifiable, since for any orthonormal matrix $\bO \in \mathbb{R}^{K \times K}$, $(\bW \bO^T) (\bO \bZ) =  \bW \bZ$. However, it does not affect how well the model fits since the linear space spanned by the columns of $\bW \bO^T$ is the same as the linear space spanned by the columns of $\bW$. We further note that for the downstream analysis using the estimated $Z$, if the downstream model maintains linearity with respect to  $Z$, its fit remains unaffected. For instance, in the context of studying the association between $Z$ and a continuous outcome variable $H$ using a linear model, we write the model for all $n$ subjects in the matrix form as $\bH = \bZ^T \tau + \bepsilon$, where $\bH, \bepsilon \in \mathbb{R}^n$ and $\tau \in \mathbb{R}^{K}$. It can be shown that $(\bO \bZ)^T (\bO \tau) =  \bZ^T \tau$ under the same orthogonal transformation matrix $\bO$ for $\bZ$. Therefore, the model fitting is essentially the same.

\subsubsection{Model fitting} \label{model_fitting}
 
We optimize the objective function in (\ref{objective}) using Adam, a state-of-the-art gradient descent learning algorithm employing adaptive learning rates for efficient optimization \citep{KingBa15}, with further details in the Supplementary Materials.
In latent variable modeling, the risk of overfitting is particularly acute with significant missing data \citep{josse2011multiple, raiko2007principal}. To address this, we implement a procedure where $20\%$ of the data is held out randomly to test for overfitting: we train on $80\%$ and monitor loss on both sets. We noted a consistent decrease in the training loss and an increased hold-out loss from certain iterations, which suggests overfitting. Hence, we employ early stopping, determining the optimal stopping point at the minimum hold-out loss, and then retrain with all data up to this point. This approach encourages the model to learn parameters that demonstrate the best generalizability on unseen data, yielding more robust and reliable results. Additionally, the selection of the tuning parameter $\lambda$ and the number of latent variables can be guided by the reconstruction error for $\{Y_{ij}\}$ observed on the hold-out sets using the model trained under the selected optimal number of iterations.

\subsection{Generalization to future population}

After training the model following the procedure introduced in Section \ref{model_estimation}, we will obtain estimates for $\bbeta := \{\beta_{jm}\}$, $\bW$, and $\bgamma := \{\gamma_j\}$, denoted as $(\widehat \bbeta, \widehat \bW, \widehat \bgamma)$, as well as the estimated $Z_i$ for each subject in the training data. For each future subject $i$ from the same institutions as the training data cohort, we are able to learn their optimal latent representation $\widetilde Z_i$ by leveraging their observed genomic alteration data $\widetilde Y_i$, institution information $\widetilde C_i$, and baseline characteristics $\widetilde V_i$. Mathematically, 
$\widetilde Z_i = \argmin_{z}
- \sum_{j \in I_{i}}
\Bigg ( \widetilde Y_{ij} \log \bigg[\sigma \Big\{ \textcolor{black}{ \sum_{m=1}^M \widehat \beta_{jm} I(\widetilde C_i = m)} +  \\
\widehat \bW_j z + \widehat \bgamma_j^T \widetilde V_{i} \Big\}\bigg]
+  (1 - \widetilde Y_{ij}) \log \bigg[ 1 - \sigma \Big \{\textcolor{black}{\sum_{m=1}^M \widehat \beta_{jm} I(\widetilde C_i = m)} + \widehat \bW_j z + \widehat \bgamma_j^T \widetilde V_{i} \Big\} \bigg] \Bigg ).  $
% \begin{dmath*}
% \widetilde Z_i = \argmin_{z}
% - \sum_{j \in I_{i}}
% \left ( \widetilde Y_{ij} \log \bigg[\sigma \Big\{ \textcolor{black}{ \sum_{m=1}^M \widehat \beta_{jm} I(\widetilde C_i = m)} + \widehat \bW_j z + \widehat \bgamma_j^T \widetilde V_{i} \Big\}\bigg] \\ 
% +  (1 - \widetilde Y_{ij}) \log \bigg[ 1 - \sigma \Big \{\textcolor{black}{\sum_{m=1}^M \widehat \beta_{jm} I(\widetilde C_i = m)} + \widehat \bW_j z + \widehat \bgamma_j^T \widetilde V_{i} \Big\} \bigg] \right ).  
% \end{dmath*}
This enables us to rectify the systematic biases stemming from sequencing technique differences estimated from the training data, while accounting for the unique characteristics of each incoming patient. 
In cases where the model is applied to a population originating from multiple institutions with a relatively large sample size $\widetilde n$, we can consider incorporating regularization terms on $\{\widetilde Z_i\}$ as well and simultaneously optimizing $\{\widetilde Z_i\}$ for all subjects to further promote alignment of $\widetilde Z$ across institutions for this new population. In that case, the objective function can be expressed as 
\begin{align*}
\min_{{\widetilde Z_i,\ \forall i}}  & - \frac{1}{\widetilde n} \sum_{i=1}^{\widetilde n} \frac{1}{|I_{i}|} \sum_{j \in I_{i}} \Bigg( \widetilde Y_{ij} \log\bigg[\sigma\Big\{\sum_{m=1}^M \widehat \beta_{jm} I(C_i = m) + \widehat \bW_j \widetilde Z_i + \widehat \bgamma_j^T \widetilde V_{i}\Big\}\bigg] \notag \\
&  + (1 - \widetilde Y_{ij}) \log\bigg[1 - \sigma\Big\{\sum_{m=1}^M \widehat \beta_{jm} I(C_i = m) + \widehat \bW_j \widetilde Z_i + \widehat \bgamma_j^T \widetilde V_{i}\Big\}\bigg] \Bigg)  \notag \\
& + \lambda \sum_{s=1}^S \sum_{k=1}^K \sum_{m=2}^M \bigg\{Q(\widetilde \bZ_{k}^{(m)}; p_s) - Q(\widetilde \bZ_{k}^{(m-1)}; p_s)\bigg\}^2.  \notag
\end{align*}

\vspace{-10pt}
\section{Simulation studies} \label{simulation}
We conducted simulation studies with two primary objectives. Firstly, we assessed the effectiveness of the latent variables $Z$ obtained through the proposed Bridge model in facilitating downstream analysis. Secondly, we evaluated the accuracy of the Bridge model parameter estimation. To achieve the first goal, we constructed an outcome of interest $O$ as $O = a^T V + b^T Z + \epsilon, \epsilon ~ \mathcal{N}(0, 5)$. We compared the predictive performance for $O$ on the test data among four different models: 
(1) ``V only", which mirrors practical scenarios where only clinical variables are used to predict patients' clinical outcomes,
(2) ``V + shared genes", which represents the traditional practice of including only genes shared across institutions in multi-institutional studies; 
(3) ``V + Z from Bridge model", where $Z$ was estimated using the proposed Bridge model, 
(4) ``V + Z from Naive Bridge", where $Z$ was estimated incorporating all available genomic data, encompassing both shared and institution-specific genes, but without accounting for the inconsistent measurements across institutions. Mathematically, $Y_{ij}$ \textcolor{black}{in the Naive Bridge approach} was modeled as $\mathbb P(Y_{ij} = 1 \mid Z_i) = \sigma (\beta_j + \bW_j Z_i)$
and no regularization on $Z$ was imposed. We included this rudimentary version of the proposed Bridge model, intended solely for the illustrative evaluation of disregarding sequencing platform variations.

\subsection{Simulation settings} \label{simulation_settings}
We considered 600 genes and two institutions, among which only a subset of genes were sequenced at both two institutions.
We employed a training sample size of $1000$, evenly distributed across the institutions.
In \textbf{Setting 1}, we simulated gene mutation status based on the model presented in (\ref{y_prob}). 
In \textbf{Setting 2}, we considered non-linear association between $Y_{ij}$ and $Z_i$ by adding a quadratic term in the likelihood so that
$
\mathbb P(Y_{ij} = 1 \mid \mathrm{C}_i = m, Z_i, V_i) = \sigma \Big(\beta_{jm} + \bW_j Z_i + \alpha \bW_j Z_i^2 + \bgamma_j^T V_{i} \Big),
$
where $Z_i^2$ denotes $[Z_{i1}^2, ..., Z_{i5}^2]^T$. Details of the simulation parameters are presented in the Supplementary Materials.
In both settings, we considered four proportions of shared genes between the two institutions: $40\%$, $20\%$, $10\%$, and $5\%$.

To evaluate the prediction accuracy of the outcome of interest $O$, in each experiment, we simulated the training and test samples separately with a fixed sample size of $500$ for the test data and calculated the rooted mean squared errors (RMSEs) of $O$ on the test set.
%Using the training data, we estimated the model parameters together with the latent $Z$ for the training samples from the proposed Bridge model and the Naive Bridge model. Additionally, we trained a linear model for each of the four comparative methods. Subsequently, these trained models were applied to the test data, and we calculated the rooted mean squared errors (RMSEs) of $O$ on the test set. This procedure was repeated $100$ times for each experiment scenario, with a fixed sample size of $500$ for the test data. 
To assess the accuracy of parameter estimates for $\{\beta_{jm}\}$, $\{\gamma_{j1}\}$, and $\{\gamma_{j2}\}$, we conducted additional experiments with sample sizes of $500$ and $1500$ under Setting 1. The mean squared error (MSE) was computed for each set of parameters. Each experiment was repeated $100$ times.

In accordance with the model training strategy proposed in Section \ref{model_fitting}, we selected the iteration number to be $3000$ to train the Bridge model and $100$ iterations to learn the latent $Z$ for the test data, under a learning rate of $0.01$. We explored different $\lambda$ values, $0.001, 0.005, 0.01$, and $0.05$, and presented the model results in the following section under the $\lambda$ value of $0.005$. To ensure a fair comparison, we also tune the number of iterations for model training and estimating $Z$ in the Naive Bridge model to avoid overfitting.
 
\subsection{Simulation results}  \label{simulation_results}
\begin{figure}
  \centering
  \subfloat{ \includegraphics[width=1 \linewidth]{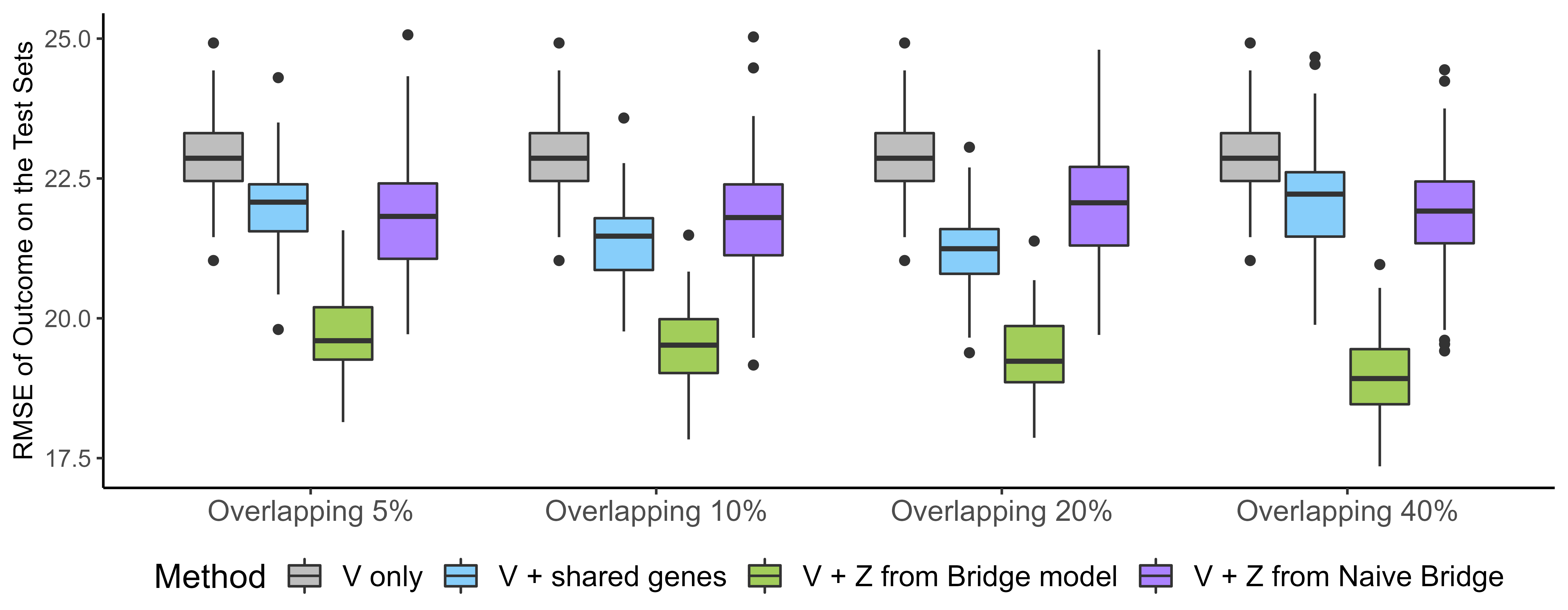}  } \\
  %\vspace{-5pt}
  \small (a) Setting 1 \\
  \vspace{-10pt} % Adjust the vertical space here
  \subfloat{ \includegraphics[width=1 \linewidth]{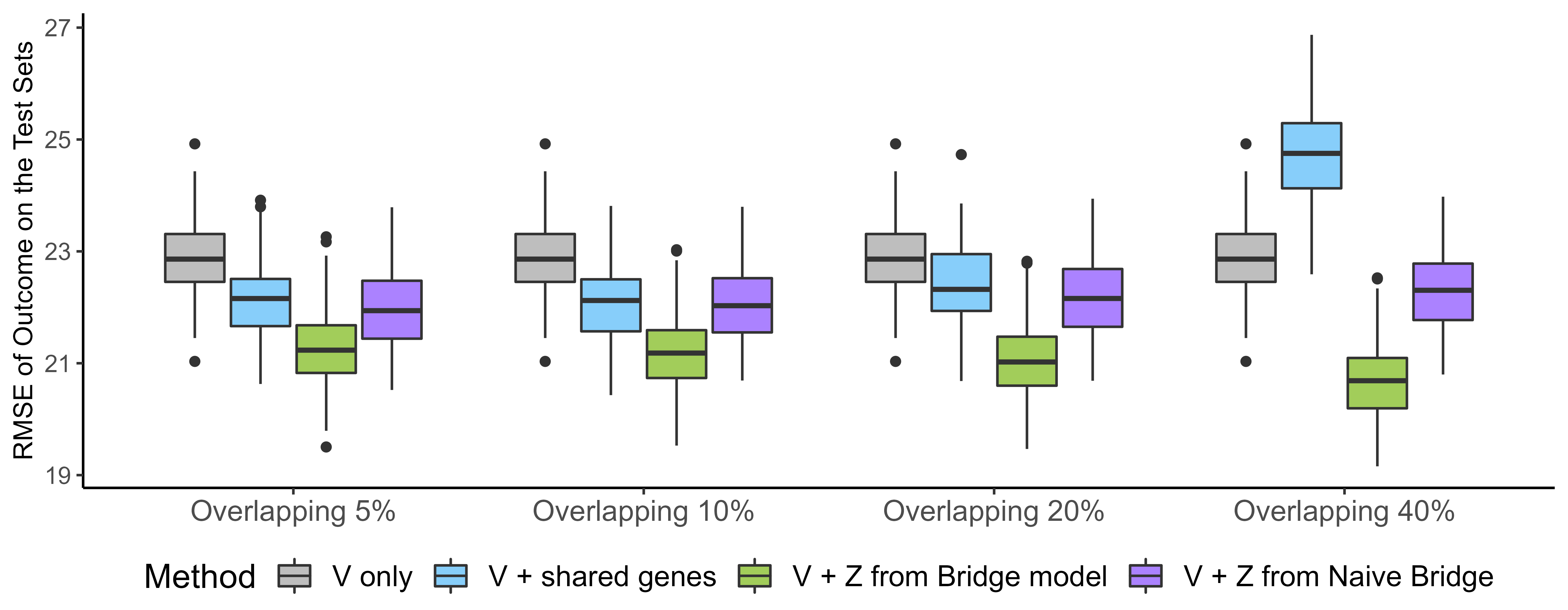} } \\
    %\vspace{-5pt}
   \small (b) Setting 2 \\
  \caption{RMSE for outcome prediction on the test datasets under different settings} \label{boxplot_rmse_outcome}
\end{figure}

The RMSEs for predicting the outcome $O$ on the test sets were presented in Figure \ref{boxplot_rmse_outcome}.

The proposed Bridge model achieves the lowest outcome prediction error on the test datasets in all settings and scenarios. Notably, it is robust to model misspecification in the presence of non-linear latent structure, as observed in Setting 2. Moreover, the model’s performance benefits from an increase in the proportion of genes that overlap across different institutions, as this provides more informative data for learning the latent variables. A similar trend is observed for the model with $V$ + shared genes, when the gene overlapping proportion increases from $5\%$ to $20\%$. However, when the gene overlap reaches $40\%$, the model’s performance declines. This is likely due to the challenge posed by the relatively large feature dimension compared to the sample size, a phenomenon often known as \textit{curse of dimensionality}. This effect is even more pronounced in Setting 2, where the addition of overlapping genes performed worse than using only the clinical variables $V$. This is primarily because overfitting can be particularly problematic in the presence of model misspecification. 

In stark contrast, the proposed Bridge model effectively mitigates the challenges posed by the \textit{curse of dimensionality} by extracting a concise set of latent features for downstream modeling. Simulation results demonstrate that it suffers much less from model misspecification, with minimal sensitivity to the proportion of overlapping genes. This highlights the reliability and suitability of the proposed model across a range of scenarios, making it particularly appealing for scenarios where genomic data lacks commonly shared genes across institutions. On the other hand, the Naive version of Bridge model performs much worse than the Bridge model, indicating the importance of addressing inconsistent mutation detections in multi-institutional studies. Simulation experiments with varying values of $b$ are presented in Supplementary Materials, where we observed robust results under Bridge model.

\begin{figure}
        \vskip -0.2 in
          \centering    
  \centering
    \includegraphics[width=1 \linewidth]{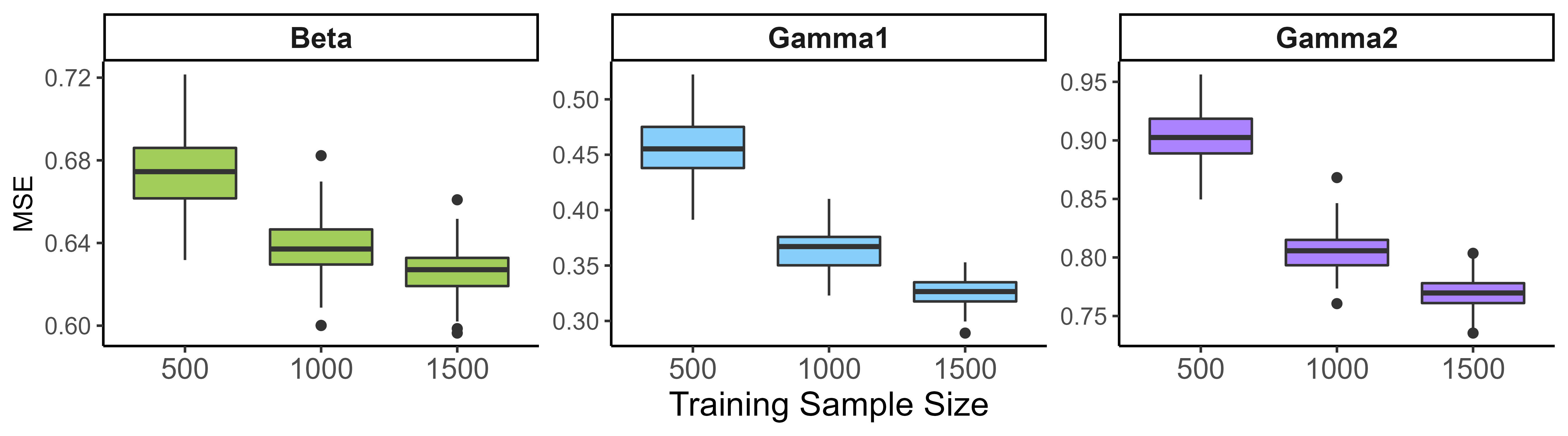}
    % \vskip -0.1 in
\vskip -0.05 in
    \caption{MSEs of each set of parameters across varying training sample sizes}
     % \vskip -0.21 in
     \label{sim_mse_param}
\end{figure}

Additionally, in Figure \ref{sim_mse_param}, we plot the MSE of the parameter estimates for $\{\beta_{jm}\}$, $\{\gamma_{j1}\}$, and $\{\gamma_{j2}\}$ respectively from 100 replications. As the training sample size increases, we observe a notable improvement in the accuracy of parameter estimation, indicating the feasibility of parameter estimation within the framework of the proposed objective function and identifiability constraints. The model training is also observed to scale effectively with sample size, with computation times detailed in the Supplementary Materials.
%estimating parameters related to the institutional differences, as well as those associated with both continuous and binary covariates in the model. These results underscore the feasibility of parameter estimation within the framework of the proposed objective function and identifiability constraints.

\section{Application to the GENIE BPC Study} \label{real_data}

To effectively leverage the multi-institutional clinico-genomic GENIE BPC data and address the challenges described in \ref{intro}, we applied the proposed Bridge model to all six cancer types in GENIE BPC: non-small cell lung cancer (NSCLC), breast cancer, prostate cancer, colorectal cancer, pancreatic cancer, and bladder cancer, where patients' data are collected from four institutions: DFCI, MSK, Princess Margaret Cancer Centre, University Health Network (UHN), and VICC. 
Our objective was to evaluate the utility of the latent variables derived from the Bridge model regarding its predictive power for overall survival (OS). To assess OS, we considered the time from metastasis to all-cause mortality using Cox proportional hazards models. We compared our approach, referred to as the ``clinical + Z from Bridge model", which includes in the Cox model the clinical variables and the latent variables extracted from the Bridge model, with two alternative Cox models: the ``clinical model", encompassing solely clinical variables, and the ``clinical + common genes" model, which incorporated clinical variables and shared genes across sequencing panels. The Cox model and Bridge model expressions, as well as more details of the model and the study population for analyses, are provided in the Supplementary Materials.

The clinical variables incorporated in our analysis included age, gender (except for breast and prostate cancer), cancer stage at diagnosis (stage IV vs. stage I-III), and histology. Histology was omitted in the Bridge model for NSCLC, pancreatic, and bladder cancer due to the complete absence of histology for UNH in these cancer types. We incorporated other clinical variables to our analysis, such as smoking status for NSCLC, subtype for breast cancer, and Gleason score for prostate cancer. These clinical variables were included in the Bridge models to account for the patient-level heterogeneity potentially associated with gene mutation variations and in the Cox models as potential prognostic factors for OS. An indicator variable for the institution was also included in every Cox model to account for this factor. For the Bridge model, \textcolor{black}{we selected $\lambda$ and the number of latent variables for each cancer type individually by evaluating the reconstruction error on intentionally held-out entries from the training data (detailed in the Supplementary Materials).} 
%we utilized ten latent variables and set $\lambda$ to $0.005$ on the quartiles of $Z$ on both the training and test data for all six cancer types for illustration. This configuration yielded favorable reconstruction errors on the hold-out validation sets during model training. 
We determined the optimal iteration number for each cancer type and for training and test data separately, as outlined in Section \ref{model_estimation}.

\begin{figure}[!t]
        \vskip -0.2 in
          \centering    
  \centering
    \includegraphics[width=1 \linewidth]{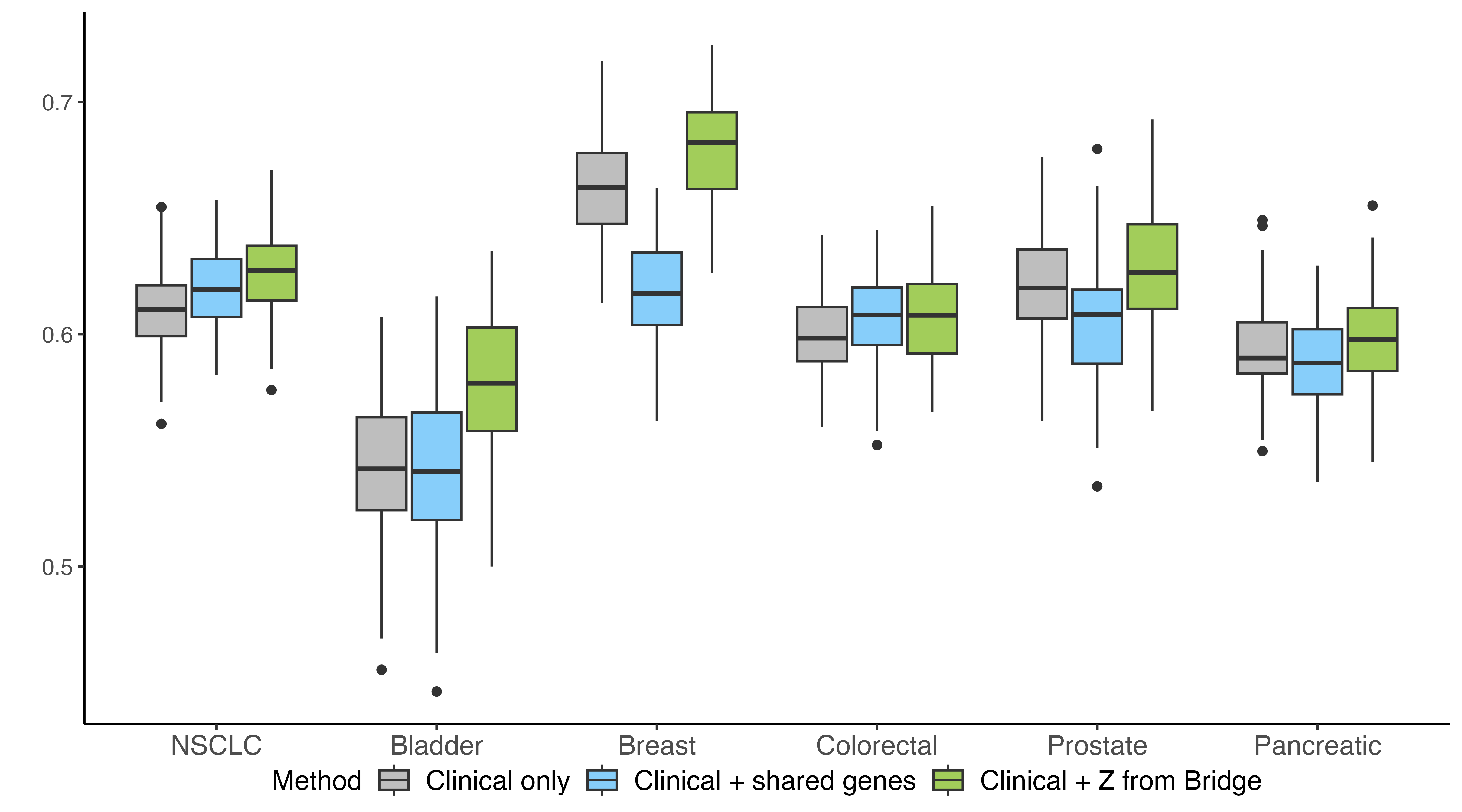}
    % \vskip -0.1 in
\vskip -0.05 in 
    \caption{\textcolor{black}{Concordance index for time to overall survival from metastasis on the test sets from 100 replications for each cancer type in GENIE BPC}}
     % \vskip -0.21 in
     \label{cox_boxplot}
\end{figure}

To assess and compare the performance of all three models, we randomly split the data into training and testing sets with a $7:3$ ratio and repeated the procedure $100$ times. In each replication, we evaluated models' prediction for OS on the test set by concordance index \citep{harrell1996multivariable}. A higher concordance index indicates a better prediction power for time-to-event outcomes. The concordance index from the 100 replications is displayed in Figure \ref{cox_boxplot}. The clinical model exhibits prognostic value for OS with a mean concordance index ranging from $0.54$ for bladder cancer to $0.66$ for breast cancer. Remarkably, the Bridge model enhances the predictive power of the clinical model across all six cancer types, with the extent of improvement varying by cancer type. The biggest improvement was observed in bladder cancer, where the mean concordance index increased from $0.54$ to $0.58$. Though a successful multi-institutional genomic data integration is not a guarantee for improved clinical outcome prediction, the results we achieved on GENIE BPC data partially reflect the utility of maximally preserving genomic information using the Bridge model. We would also like to note that the enhancement in prediction accuracy from genomic features is cancer-type dependent, as not all types hold the same predictive relevance for survival outcomes.
%clinical relevance of the latent variables we've extracted from the comprehensive amalgamation of individual mutation data. 

In contrast, for all cancer types except NSCLC and colorectal cancer, the inclusion of common genes did not significantly enhance, and in some cases, compromised, the performance of the clinical model. This could be due to the loss of information when institution-specific genes are ignored, along with the introduction of noise introduced from individual genes. This noise may arise from the differences in sequencing methods and techniques across institutions, resulting in inconsistent measurements that obscure true signals, or from individual genes that hold little to no prognostic value for OS. However, across all cancer types, the proposed Bridge model consistently outperformed the model based on common genes alone, highlighting the importance of aggregating comprehensive genomic data while effectively mitigating the variabilities at the institutional level.
%Our approach thus unveils a richer set of clinically relevant features, which enhances the model's overall predictive capacity and generalizability.

Additionally, we examined the estimated latent variables and parameters in the Bridge model. The relationship between the latent variables and the gene mutations can be investigated based on the fitted $\bW$ in the Bridge model. We illustrate this with heatmaps of the top 20 gene mutations for each latent variable and interpret their association with overall survival using fitted Cox model coefficients, all of which are detailed in the Supplementary Materials. Furthermore, we examined the fitted ${\beta_{jm}}$ parameters in the Bridge model, which account for the disparities in mutation detection attributable to sequencing platform differences across institutions. We observed good concordance between the relative differences in ${\beta_{jm}}$ and the observed mutation differences across institutions (detailed in the Supplementary Materials), indicating that the observed mutation differences across institutions can be largely attributed to technical variations, while some variability is due to true patient heterogeneity.

\section{Discussion} \label{discussion}
In this paper, we introduce the Bridge model, specifically designed to harmoniously integrate high-dimensional genomic data from diverse institutions. It addresses missing data and systematic biases due to differences in the sequencing panel and assay techniques while preserving true biological variations among individuals. By fully utilizing available genomic data, it enables a comprehensive characterization of each individual's genomic profile.

This method is motivated by the pressing real-world challenge posed by multi-institutional genomic studies, which are gaining increasing significance as genomic sequencing becomes standard practice for various cancer types \citep{chakravarty2021clinical}. Addressing these complexities and challenges presents a unique opportunity to harness and leverage such powerful real-world evidence. It paves the way for more comprehensive and accurate analyses, enhancing our ability to gain critical insights from the wealth of genomic data and ultimately advancing the field of precision oncology.

We envision several promising avenues for future research. First, regularization terms can be explored for the Bridge model parameters, like L1 for sparsity and L2 to prevent overfitting. Additionally, expanding the model to include more complex latent structures could enable the capture of non-linear associations, potentially enhancing the capacity and informativeness of the latent variables in explaining genomic variations. Demonstrating the model's effectiveness in clinical outcome prediction with GENIE BPC data opens up the potential for exploring its use in distinguishing treatment effects, offering a step toward personalized treatment recommendations and advancing precision oncology. \textcolor{black}{Recent work by \cite{wang2024multiple} introduced a multiple augmented reduced rank regression approach, which accounts for both individual and global cancer effects for pan-cancer analyses. Our model could also be potentially extended for integrative pan-cancer analyses by leveraging cross-cancer strengths while accommodating unique characteristics of different cancer types.}

\backmatter

%  This section is optional.  Here is where you will want to cite
%  grants, people who helped with the paper, etc.  But keep it short!

\section*{Acknowledgements}

The authors thank Jessica Lavery, Samantha Brown, and Hannah Fuchs for their help with the AACR GENIE BPC data preparation. This research is supported by NIH grants 3P50CA271357-02S1, P30-CA008748, R25CA272282, and the MSK Society.

\textcolor{black}{\section*{Supplementary Materials}
Web Appendices referenced in Sections \ref{model_fitting}, \ref{simulation_settings}, \ref{simulation_results}, and \ref{real_data}, along with the code, are available with this paper at the Biometrics website on Oxford Academic. Code for the proposed Bridge model and for running the simulation studies is available on the GitHub repository at \\ https://github.com/ychen178/Bridge\_model.
\vspace*{-8pt}}

\section*{Data Availability}
The data that support the findings \textcolor{black}{in this paper are available from the American Association for Cancer Research (AACR) Project Genomics Evidence Neoplasia Information Exchange (GENIE) Biopharma Collaborative (BPC).} Specifically, data pertaining to non-small cell lung cancer and colorectal cancer are already accessible to the public \textcolor{black}{at \\
https://www.aacr.org/professionals/research/aacr-project-genie/bpc}. Data for the other cancer types will become publicly available and are currently subject to approval by AACR Project GENIE.

\bibliographystyle{biom} 
\bibliography{refs_genie} %refs_deepnext

%\newpage

%\begin{appendices}

% \renewcommand\thefigure{\theappendix.\arabic{figure}}
% \numberwithin{figure}{section}
% \section{Appendix}

%\end{appendices}

\end{document}